\documentclass[12pt,preprint]{aastex}

\usepackage{natbib}

\newcommand{\simgt}{\lower.5ex\hbox{$\; \buildrel > \over \sim \;$}}
\newcommand{\simlt}{\lower.5ex\hbox{$\; \buildrel < \over \sim \;$}}
\newcommand{\phisma}{\phi_{\mbox{{\tiny SMA}}}}
\newcommand{\phitp}{\phi_{\mbox{{\tiny TP}}}}

\newcommand{\thetael}{\theta_{\mbox{\tiny{EL}}}}

\shorttitle{Atmospheric Phase Correction at the Submillimeter Array}
\shortauthors{Battat et al.}

\begin{document}


\title{Atmospheric Phase Correction Using Total Power Radiometry \\at the Submillimeter Array}


\author{James B.~Battat\altaffilmark{1,2}}
\author{Raymond Blundell\altaffilmark{1}}
\author{James~M.~Moran\altaffilmark{1}}
\author{Scott Paine\altaffilmark{1}}
\affil{Harvard-Smithsonian Center for Astrophysics, Cambridge, MA 02138}


\altaffiltext{1}{Harvard-Smithsonian Center for Astrophysics, 60 Garden Street, MS-10, Cambridge, MA 02138}
\altaffiltext{2}{jbattat@cfa.harvard.edu}


\begin{abstract}
Phase noise caused by an inhomogeneous, time-variable water vapor distribution in our atmosphere reduces the angular resolution, visibility amplitude and coherence time of millimeter and submillimeter wavelength interferometers.
We present early results from our total power radiometry phase correction experiment carried out with the Submillimeter Array on Mauna Kea.  
From accurate measurements of the atmospheric emission along the lines of sight of two elements of the array, we estimated the differential atmospheric electrical path between them.
%
%
%
In one test, presented here, the phase correction technique reduced the rms phase noise at 230 GHz from 72$\degr$ to 27$\degr$ over a 20 minute period with a 2.5 second integration time.  This corresponds to a residual differential electrical path of 98 $\mu$m, or 15 $\mu$m of precipitable water vapor, and raises the coherence in the 20 minute period from 0.45 to 0.9.  
\end{abstract}


\section{Introduction}
The water vapor in our atmosphere is spatially inhomogeneous and highly time-variable, resulting in atmospheric electrical path fluctuations that alter the phase of propagating electromagnetic waves.
At submillimeter wavelengths, the effect of atmospheric fluctuations is particularly severe, and limits the resolution and coherence of interferometric arrays.
Without atmospheric phase correction schemes, submillimeter wave interferometers such as the Submillimeter Array\footnote{The Submillimeter Array is a joint project between the Smithsonian Astrophysical Observatory and the Academia Sinica Institute of Astronomy and Astrophysics, and is funded by the Smithsonian Institution and the Academia Sinica.  For more on the SMA and its characteristics see \citet{ho_moran_lo}.} (SMA) and the upcoming Atacama Large Millimeter Array (ALMA)
will not be able to image at their diffraction limits in cases where self-calibration or fast switching is not possible.  The advantages and limitations of fast switching are described by both \citet{carilli_holdaway} and \citet{layb}.  At the SMA, fast switching may have limited value because of the large angular distances between calibrator sources of sufficient strength \citep{battatSMA}.

The atmospheric phase at millimeter and submillimeter wavelengths can be determined empirically by monitoring the continuum emission at the same frequency as the astronomical observation.  
The use of a single receiver for both source observation and phase calibration makes it possible to monitor the volume of atmosphere responsible for the phase errors.
The accuracy of the phase correction is then independent of baseline length.
Such a phase correction scheme can not only improve the quality of observations, but also broaden the range of weather conditions during which observations are possible.  This translates to a more efficient observing schedule.

We have developed highly stable total power detectors and have explored how reliably they can be used for accurate phase correction.
The initial results described here show that total power radiometry has great promise to solve the problem of image blurring caused by atmospherically induced phase errors.

\section{Atmospheric Phase Errors}
\label{atmosphere}
Atmospherically induced phase errors degrade the measured visibility amplitudes, image resolution and dynamic range of submillimeter interferometers \citep[e.g.][]{carilli_holdaway}.
Although water vapor is essentially non-dispersive at radio frequencies (one mm of precipitable water vapor, {\em pwv}, creates $\sim$6.5 mm of excess electrical path), the corresponding phase errors increase linearly with frequency.  
In the immediate vicinity of strong atmospheric lines, the water vapor is dispersive, but at these frequencies the high opacity makes ground based radio astronomy impossible \citep{sutton_hueckstaedt}.
In the well established Kolmogorov description of atmospheric turbulence, the root phase structure function, which describes how phase errors depend on the antenna separation or baseline, $b$, is a broken power law proportional to $b^{\beta/2}$, where $\beta/2$ is called the Kolmogorov exponent.  
For baselines shorter than the thickness of the turbulent layer ($b \simlt 0.5-2$ km), the turbulence is approximately isotropic in three dimensions and $\beta/2 = 5/6$.  
For baselines longer than the turbulent layer thickness, the turbulence is two dimensional and $\beta/2 = 1/3$ \citep{carilli_holdaway}.
On size scales larger than the outer scale of atmospheric water vapor (typically 5-10 km), the phase errors are independent of baseline length.  
Because of this outer scale and the associated long time scales, Very Long Baseline Interferometry ($b > 1000$ km) is viable.

The expected rms phase error between antennas pointed to the zenith can be written as
	\begin{equation}
	\label{phiRMS}
	\sigma_\phi = K \left(\frac{\nu}{\mbox{230 GHz}}\right) \left(\frac{b}{\mbox{100 m}}\right)^{\beta/2}\csc^\gamma\left(\thetael\right),
	\end{equation}
where $\nu$ is the observing frequency, $\thetael$ is the elevation angle of the observation and $\gamma$ is 1/2 or 1 for three or two dimensional turbulence, respectively.
Measurements at the SMA site, in the three dimensional turbulence regime, give $K\sim 30\degr$ and $\beta/2\sim 0.75$, though large variations in both parameters are seen \citep{masson}.
If the distribution of phase fluctuations is Gaussian, then the expectation of the measured visibility amplitude, $\left<V_m\right>$ (where $\left<\ldots\right>$ denotes an ensemble average) is related to the true visibility amplitude, $V$, by
	\begin{equation}
	\label{visibility}
		\frac{\left<V_m\right>}{V} = e^{-\sigma^2_\phi/2}
	\end{equation}
\citep{tms}.  
Thus an rms phase error of 1 radian, which is expected on a 140 m baseline zenith observation at 345 GHz (Equation \ref{phiRMS}), introduces a 40\% coherence loss.
The resolution of the astronomical image is degraded by a convolution of the true image with a seeing function that depends on the root phase structure function \citep{tms}.  
This effect is the submillimeter analog to optical seeing.
\citet{masson} showed that the median resolution allowed by the Mauna Kea atmosphere is 1\farcs2 at 345 GHz, corresponding to an effective maximum baseline of about 100 m.


The dominant contribution to fluctuations in atmospheric delay comes from poorly mixed, 
wind-blown water vapor cells of various sizes that constitute a turbulent region confined to the first few kilometers above ground level \citep{woody}.
The relevant time scale for these fluctuations is $\sim b/v$ where $v$ is the wind speed along the baseline.
At the SMA, $b$ is 10 to 500 m and $v$ is typically 5 to 10 m s$^{-1}$, so that the dominant fluctuations have periods of 1 to 100 seconds.

Because surface meteorological measurements cannot provide an accurate determination of total water vapor content \citep{waters}, remote sensing measurements, such as microwave radiometry, are necessary.
The real and imaginary parts of the dielectric constant of water vapor are related by the Kramers-Kronig relation and thus there is a strong correlation between atmospheric opacity and phase delay \citep{tms}.
The opacity in the atmospheric continuum windows below 1 THz is dominated by the far wings of infrared water resonance lines and also includes a continuum component that is proportional to the square of frequency.
The correlation between the atmospheric brightness temperature and water vapor content has been established empirically by comparing radiometer measurements with radiosonde data \citep[e.g.][]{staelin,westwater_guiraud,moran_rosen,
guiraud}.  


\section{Past Results}
The correlation between atmospheric phase delay and atmospheric brightness temperature has been measured at frequencies near water vapor resonance lines and in the continuum between the lines \citep[e.g.][]{welch}.


The Owens Valley Radio Observatory (OVRO), using 22 GHz water line radiometers, has demonstrated a reduction in the rms phase error over 25 minutes at 100 GHz from 64 degrees to 19 degrees, corresponding to a residual excess electrical path of 160 $\mu$m \citep{woody}, and have shown that phase correction improved their astronomical image \citep{marvel}.  \citet{wiedner} used a 183 GHz water line monitor to correct for atmospheric phase errors in an observation at 354 GHz with the Caltech Submillimeter Observatory (CSO) and James Clerk Maxwell Telescope (JCMT) as a two element interferometer of baseline 164 m.  The rms phase over 30 minutes was improved from 60 to 26 degrees (61 $\mu$m of residual excess path).

Total power phase correction at the frequency of astronomical observation, which requires exceptional receiver gain stability of a few parts in 10,000, has also achieved some success.  
\citet{zivanovic} used this technique at the Berkeley-Illinois-Maryland-Association Millimeter Array (BIMA). 
The total power difference at 88 GHz 
correlated with slow drifts in the measured phase but there was significant disagreement on short ($\sim$30 second) and long ($\sim$30 minute) time scales.
\citet{bremer_guilloteau_lucas} at the Institut de Radio Astronomie Millim\'{e}trique Plateau de Bure Interferometer (IRAM, PdBI) reported a consistent correlation between the total power difference, measured at 90 and 230 GHz, and the interferometric phase at those frequencies, under clear sky conditions.  They improved the rms of phase data from $>100$ degrees to 25-35 degrees at 230 GHz over a 1 minute time scale (100 $\mu$m of path).
In the presence of clouds, the total power correction technique is seriously compromised because of liquid and frozen water \citep{waters}.


\section{Total Power Radiometry for Atmospheric\\Phase Correction at the SMA}
\label{receiver}
The SMA is an interferometric array on Mauna Kea with baselines up to 500 meters and frequency coverage from 180 GHz to 900 GHz.  At the time of this experiment, integration times, $t$, as small as 2.5 seconds were possible.  
Gain variations, caused by cryocooler temperature fluctuations, typically limit the Superconductor Insulator Superconductor (SIS) heterodyne receiver \citep{blundell} stability to 0.1-1\%.

On Mauna Kea, the median rms differential path on a 100 m baseline is about 112 $\mu$m \citep{masson}, which corresponds to atmospheric brightness temperature fluctuations of 0.25 K at 230 GHz.  
Typical system temperatures at 230 GHz at the SMA are 125 K, so the atmospheric temperature fluctuations are 500 times less than the system noise temperature.
Thus a receiver stability of order 1 part in 5,000 is desired to ensure accurate measurement of atmospheric temperature fluctuations.


A servo system that monitors the physical temperature of the mixer and adjusts the receiver gain to stabilize the receiver output has been developed and tested \citep{battat}.
Our two matched total power detectors operate near room temperature and are temperature stabilized to 1 part in 10$^5$.  
The rms stability of the continuum detector, with a 33 ms integration time, was measured at 1 part in 8,000 over 10 minutes, which is only 15\% greater than the fundamental $1/\sqrt{Bt}$ thermal noise.
The servo system was inactive during the tests described in this paper.  
Instead we relied on the intrinsic stability of the two cryocoolers which are known to be very stable.  The stability of the entire receiving system was measured at 5 to 7 parts in 10$^4$ on a fixed temperature load. 

Systematic instrumental noise due to radiometer gain fluctuations translates to a noise limit in the measurement of the differential electrical path between antennas, $\Delta L$, that depends on the system noise temperature, $T_{sys}$, and atmospheric conditions in the following way
	\begin{equation}
\label{deltaL}
	\Delta L \approx 35
		\left(\frac{dL/dpwv}{6.5}\right)
		\left(\frac{dT_{bri}/dpwv}{16\mbox{ K mm$^{-1}$}}\right)^{-1}
		\left(\frac{T_{sys}}{125\mbox{ K}}\right)
		\left(\frac{\Delta G/G}{5\cdot10^{-4}}\right)
		\mbox{ $\mu$m}
	\end{equation}
where $\Delta G/G$ is the fractional receiver stability and $T_{bri}$ is the atmospheric brightness temperature which is approximately $\tau_0 T_0\csc\left(\thetael\right)$ where $\tau_0$ is the zenith opacity and $T_0$ is the physical temperature of the atmosphere.  Also, $T_{sys} \approx T_{rx} + T_{bri}$ where $T_{rx}$ is the receiver noise temperature.
The value of $dT_{bri}/dpwv$ is computed from a mean atmospheric model for Mauna Kea in January \citep{paine}.  At 230 GHz it is largely independent of water vapor content, ranging from 16.4 to 15.8 K mm$^{-1}$ for 1 to 4 mm {\em pwv}.
Thus for $\Delta G/G = 1/2000$ in median 230 GHz observing conditions ($dL/dpwv\sim 6.5$, $dT_{bri}/dpwv\sim16 $ K mm$^{-1}$, $T_{sys} \sim$ 125 K) the systematic instrumental noise is 35 $\mu$m, or 10 degrees of phase, for zenith observations.  The thermal noise in the detector contributes only 1 $\mu$m path error in 2.5 seconds of integration.


\section{Observations and Results}
\label{results}
%
\begin{figure}[t]
\plotone{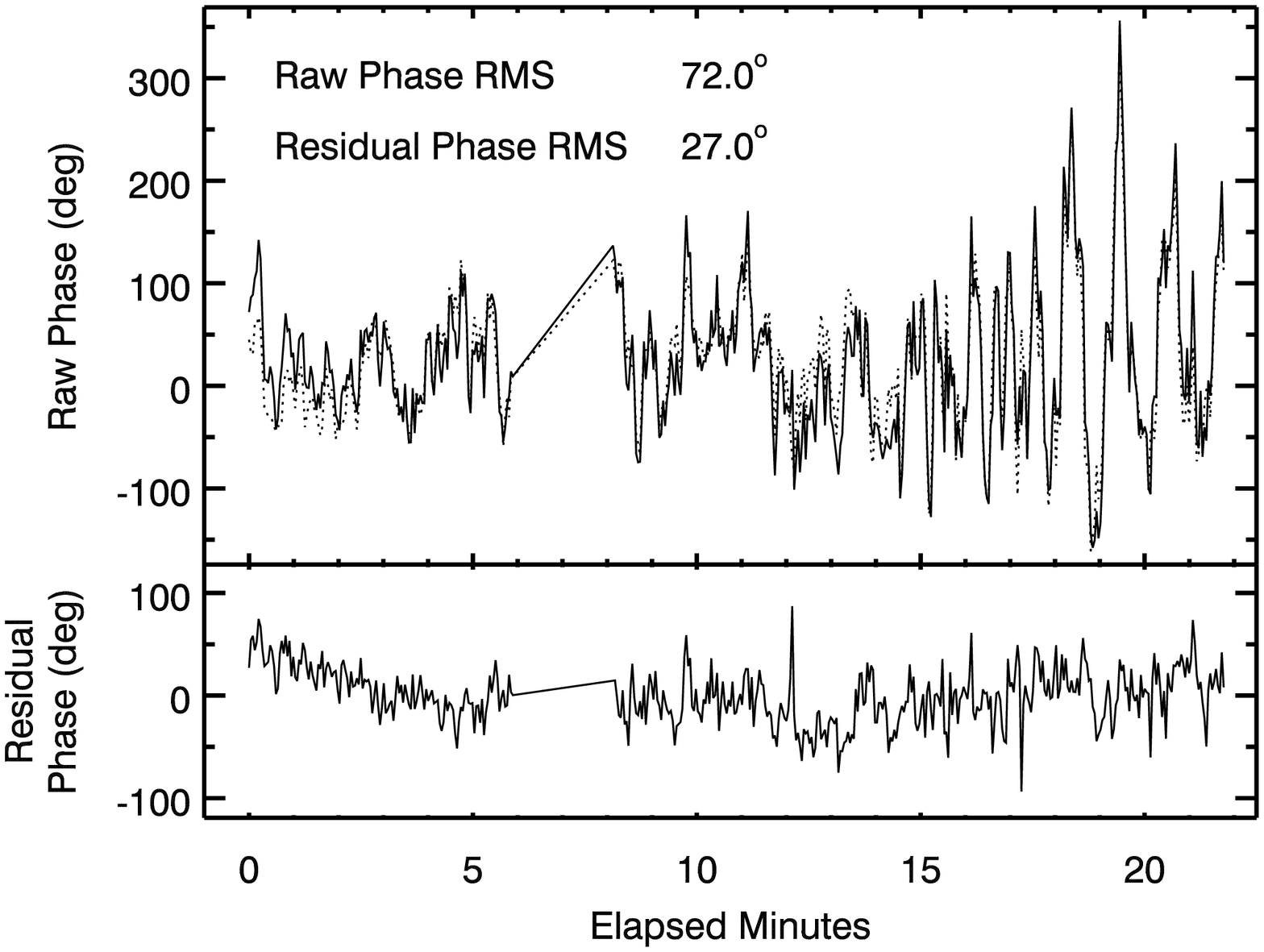}
\caption{Phase correction data from 2004 Jan 30 tracking the quasar 3C273 for 20 minutes at 230.5 GHz.   {\em Top}:  The solid line shows $\phisma$, the phase as measured by the SMA with a 2.5 second integration time, while the broken line shows $\phitp$, the phase determined from the total power measurements at the same frequency as the science observations.  The opacity increased sharply after the 15 minute mark.  {\em Bottom:}  The residual phase, $\phisma\!\!-\phitp$.}
\label{phase}
\end{figure}

We made test astronomical observations during several nights in January, 2004 with various integration times, observing frequencies and sources.  Here we present 20 minutes of data on the quasar 3C273
at 230.5 GHz with two antennas on a 140 m north-south baseline from 16:35 UTC to 16:55 UTC with a 2.5 second integration time.  
Over this time the source azimuth and elevation spanned 246.7-250.7 and 51.7-47.0 degrees, respectively.  
The surface air temperature ranged from -5.1 to -4.7 $\degr$C, and the wind speed ranged from 5 to 15 m s$^{-1}$ 
coming from the east and east-south-east.
The 225 GHz zenith opacity was between 0.13 and 0.14 during the first part of the observation and then rapidly increased to 0.18 15 minutes into the observation.
This opacity change was accompanied by the onset of large amplitude phase fluctuations (see Figure \ref{phase}).
Note that increased turbulence is not necessarily associated with increased total opacity \citep[see e.g.][]{hinder}.
The opacities listed above correspond to zenith {\em pwv} content between 3 and 4.2 mm, which is considered poor weather (75th percentile) on Mauna Kea \citep{masson}.
At 1 part in 25, the rms to mean fluctuations in the individual total power signals were easily detectable above our 5 parts in 10$^4$ noise floor.  


\begin{figure}[t]
\plotone{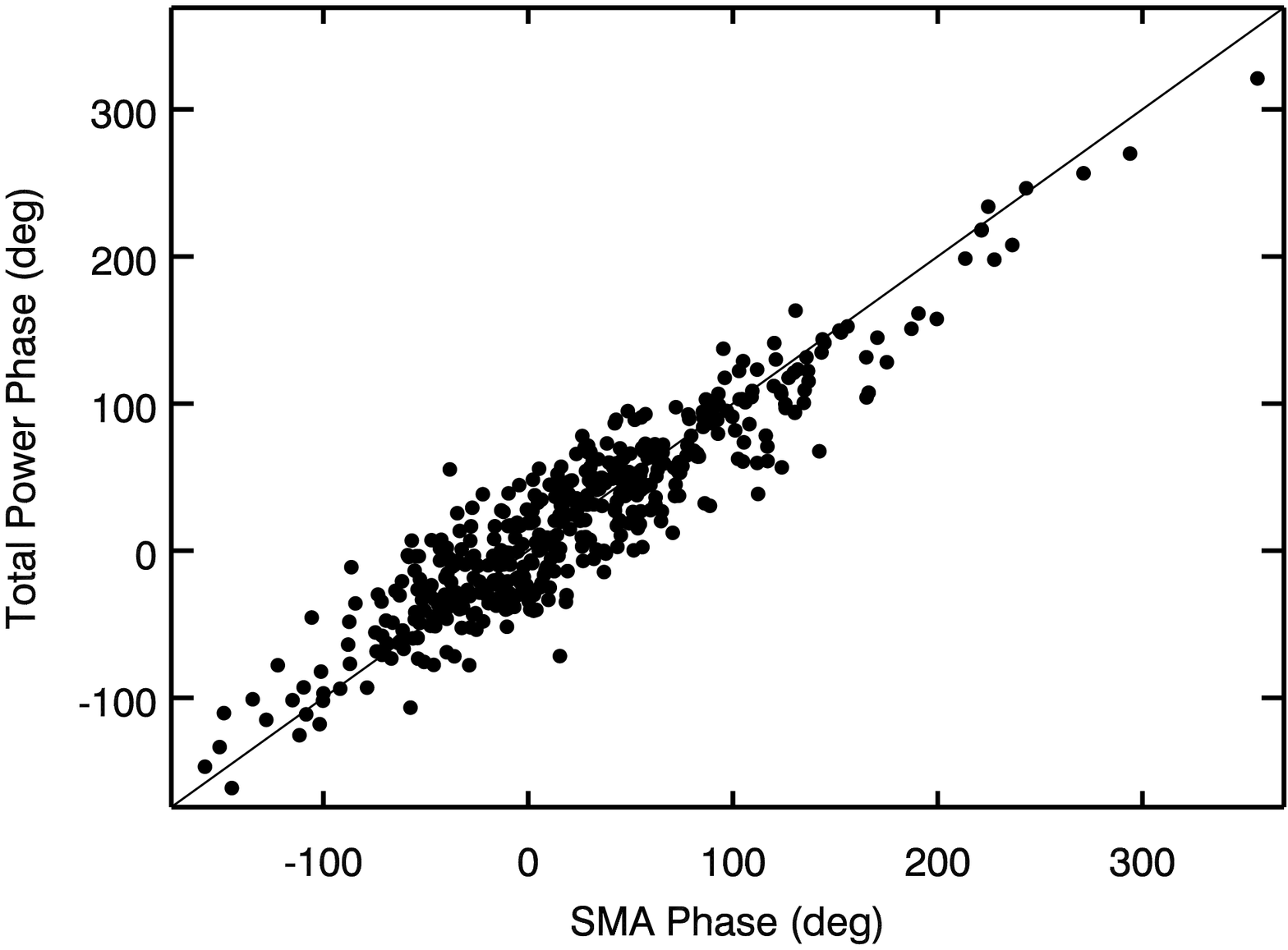}
\caption{Correlation between the interferometer phase, $\phisma$, (abscissa) and the total power phase, $\phitp$, as given by Equation \ref{tpPhase} (ordinate).  The solid line has unity slope and zero intercept.  
}
\label{correlation}
\end{figure}

\begin{figure}[t]
\plotone{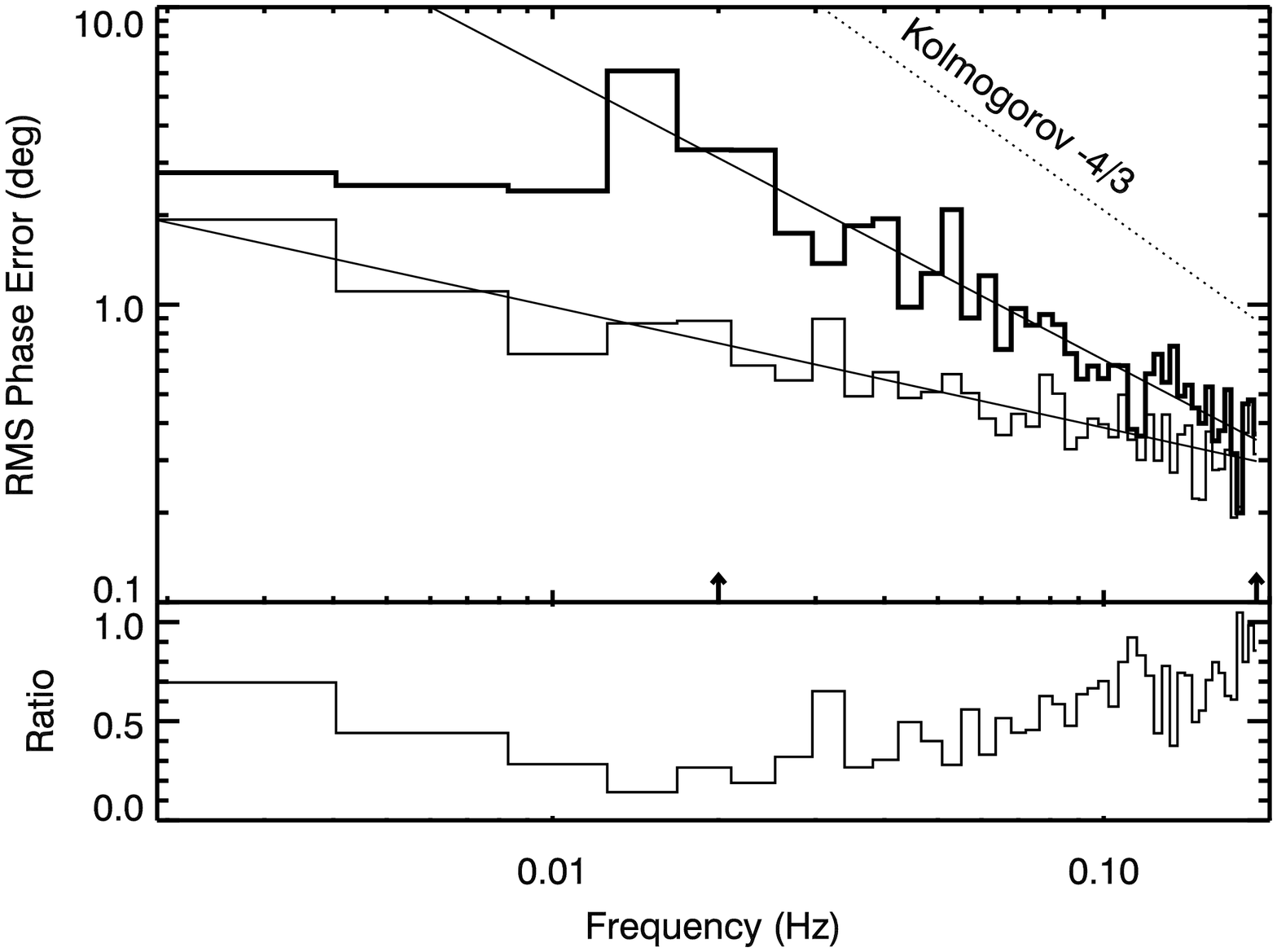}
\caption{{\em Top:} Root phase power spectrum of $\phisma$ (thick line) and $\phisma\!\!-\phitp$ (thin line).  Also shown are the Kolmogorov prediction ($\alpha = 4/3$, dotted line) and the best fits to the raw and residual data between frequencies marked with arrows ($\alpha=1.0$ and $\alpha=0.41$, respectively).  {\em Bottom:} The ratio of the residual phase and $\phisma$ plotted on log-linear axes.}
\label{power}
\end{figure}

The top panel in Figure \ref{phase} shows $\phisma$, the interferometer phase as measured by the SMA correlator and $\phitp$ the total power phase prediction which was computed from the total power measurements made at two antennas, $P_1$ and $P_2$, by the equation
\begin{equation}
\label{tpPhase}
	\phitp \equiv\left(A P_1 - B P_2\right) + C.
\end{equation}
The scaling factors, $A$ and $B$, were required because we did not have accurate temperature scales for either $P_1$ or $P_2$, and $C$ was required because of the $2\pi$ ambiguity of the interferometer phase, $\phisma$.  
To compare $\phisma$ with $\phitp$, we binned the total power data, sampled at 30 Hz, onto the 2.5 second SMA integrations.
We determined $A$, $B$ and $C$ by minimizing the mean squared error of $\phisma-\phitp$.
We did not remove any temporal drifts from $P_1$, $P_2$ or $\phisma$.


Subtracting the total power phase model from $\phisma$ improves the rms error from 72$\degr$ to 27$\degr$, corresponding to a residual differential electrical path of 98 $\mu$m or a differential water column of 15 $\mu$m.  
Figure \ref{correlation} shows the strong correlation between $\phitp$ and $\phisma$.
The coherence factor (see Equation \ref{visibility}) is improved from 0.45 to 0.90 by the total power phase correction.
During the observation, the source elevation was $\sim$50$\degr$, thus the noise floor was 11$\degr$ (see Equation \ref{deltaL}).
The expected thermal phase noise of the SMA for 3C273 (with a flux of 4 Jy at $\lambda = 0.9$ mm) was about 5$\degr$.  These independent non-atmospheric phase errors add in quadrature to contribute $\sim$12$\degr$ to the residual phase.  The remaining error in phase is attributed, in part, to unmodeled time variations in the $A$, $B$ and $C$ parameters and the lack of perfect correlation between total power and phase.


The root temporal power spectrum, $\phi(f)$, describes the rms phase error as a function of frequency, $f$.  Kolmogorov theory, under the assumption that the turbulent velocity is small compared with the wind speed (frozen turbulence, or the Taylor hypothesis), predicts $\phi(f) \propto f^{-\alpha}$ with $\alpha = \beta/2 + 0.5$.  Thus $\alpha=4/3$ above the corner frequency, $f_c\sim v/b$, and $\alpha-1=1/3$ below $f_c$ \citep{masson}.  
The root temporal power spectrum of our observation is shown in Figure \ref{power}.  The best fit function has $\alpha = 1.0$ (slightly shallower than the Kolmogorov prediction) with a corner frequency of $\sim$ 0.02 Hz.  This corresponds to a time scale of 50 seconds or a wind speed along the baseline of $\sim$ 3 m s$^{-1}$, consistent with surface meteorology.  This type of spectral break has been reported by others \citep[e.g.][]{laya,masson}.

The root temporal power spectrum of the residual phase has $\alpha=0.41$, significantly flatter than the raw phase spectrum.  
With data from a single baseline, it is difficult to determine the residual root phase structure function from the residual temporal power spectrum in order to extrapolate our results to longer baselines.  However, a worst case scenario would be to assume that the phase errors continued to grow with baseline as $\beta/2=0.75$ as is typical for uncorrected measurements on Mauna Kea.  Then diffraction limited imaging would be possible on baselines as large as 380 m.  It is more reasonable that $\beta/2 \leq 0.6$, in which case diffraction limited imaging would be possible on all SMA baselines.  

\section{Discussion and Conclusions}
The data presented above demonstrate that total power radiometry at the SMA can provide an accurate measure of the atmospheric phase between antennas and can help recover diffraction-limited image resolution.
The structure seen in the residual phase (27$\degr$ rms) is partly caused by instrumental noise from the total power radiometers and interferometer (see Section \ref{results}) and could also be due to changes in the vertical profile of the water vapor, or the presence of liquid water which contributes significantly to the atmospheric brightness temperature but not to the electrical path \citep{carilli_holdaway}.  In fact, when the data set was divided into two 10 minute segments and $A,B,C$ were derived for each segment, the resulting rms of the entire observation was improved to 22$\degr$ rms, or 83 $\mu$m of electrical path, which suggests that these parameters are sensitive to changing meteorological conditions.

In the future, the total power phase correction could be applied in real time by rapidly adjusting the phase of the local oscillator in each antenna,
or it could be done during data analysis.
In any case, the $A$, $B$ and $C$ coefficients in Equation \ref{tpPhase} should be empirically determined from measurements of a phase calibration source and then applied to the science source observations that immediately follow.  
To achieve this goal, further studies must be undertaken to determine how rapidly $A$ and $B$ vary in time, and over what range of elevations they are valid before changes in contaminating factors such as ground pickup become significant. 
In addition, when multiple baselines are present in an observation, it will be possible to derive the $A$ and $B$ parameters from a simultaneous fit to the data from all baselines.  This approach would provide a more robust parameter determination.


The development of a consistently accurate system to remove atmospheric phase delay is of immediate importance to the development of submillimeter wave interferometry.  At present no such system exists, but our data show that total power radiometry may be a viable option and is worthy of further development.

\acknowledgments
We thank R. Christensen, T. R. Hunter, R. Kimberk, P. S. Leiker and M. Wiedner for their help with instrumentation.  J.B. was supported by the NDSEG graduate fellowship.





\clearpage






























\begin{thebibliography}{}
\bibitem[Battat(2004)]{battatSMA} Battat, J.~B. 2004, SMA Technical Memo, in progress
\bibitem[Battat et al.(2004)]{battat} Battat, J.~B. et al. 2004, IEEE Trans. Microwave Theory Tech., accepted
\bibitem[Blundell et al.(1995)]{blundell} Blundell, R. et al. 1995, IEEE Trans. Microwave Theory Tech., 43, 933
\bibitem[Bremer et al.(1996)]{bremer_guilloteau_lucas} Bremer, M., Guilloteau, S., \& Lucas, R. 1996, in ESO-IRAM-NFRA Onsala Workshop, Science with Large Millimetre Arrays, ed P. Shaver (New York: Springer), 371
\bibitem[Carilli, \& Holdaway(1999)]{carilli_holdaway} Carilli, C.~L., \& Holdaway, M.~A. 1999, Radio Sci, 34, 817
\bibitem[Guiraud et al.(1979)]{guiraud} Guiraud, F.~O., Howard, J., \& Hogg, D.~C. 1979, IEEE Trans. Geosci. Electron., 17, 129
\bibitem[Hinder(1972)]{hinder} Hinder, R.~A. 1972, J. Atmos. Terr. Phys., 34, 1171
\bibitem[Ho et al.(2004)]{ho_moran_lo} Ho, P.~T.~P., Moran, J.~M., \& Lo, K.~Y. this volume, 1
\bibitem[Lay(1997a)]{laya} Lay, 0.~P. 1997a, A\&AS, 122, 535
\bibitem[Lay(1997b)]{layb} Lay, 0.~P. 1997b, A\&AS, 122, 547
\bibitem[Marvel, \& Woody(1998)]{marvel} Marvel, K.~B., \& Woody, D.~P. 1998, Proc. SPIE, 3357, 442
\bibitem[Masson(1994)]{masson} Masson, C.~R. 1994, in ASP Conf. Ser. 59, Astronomy with Millimeter and Submillimeter Wave Interferometry, ed M. Ishiguro, \& Wm.~J. Welch (San Francisco: ASP), 87
\bibitem[Moran, \& Rosen(1981)]{moran_rosen} Moran, J.~M., \& Rosen, B.~R. 1981, Radio Sci, 16, 235
\bibitem[Paine(2004)]{paine} Paine, S. 2004, SMA Technical Memo \#152
\bibitem[Staelin(1966)]{staelin} Staelin, D.~H. 1966, J Geophys Res, 71, 2875
\bibitem[Sutton, \& Hueckstaedt(1996)]{sutton_hueckstaedt} Sutton, E.~C., \& Hueckstaedt, R.~M. 1996, A\&A, 119, 559
\bibitem[Thompson et al.(2001)]{tms} Thompson, A.~R., Moran, J.~M., \& Swenson, G.~W. Jr. 2001, Interferometry and Synthesis in Radio Astronomy, (2d ed.; New York: Wiley)
\bibitem[Waters(1976)]{waters} Waters, J.~W. 1976 in Methods of Experimental Physics Vol. 12b, ed. M.~L. Meeks (New York: Academic Press), 142
\bibitem[Welch(1999)]{welch} Welch, Wm.~J. 1999, in Review of Radio Science 1996-1999, ed. W.~R. Stone (Oxford: Oxford University Press), 787
\bibitem[Westwater, \& Guiraud(1980)]{westwater_guiraud} Westwater, E.~R., \& Guiraud, F.~O. 1980, Radio Sci, 15, 947
\bibitem[Wiedner et al.(2001)]{wiedner} Wiedner, M.~C., Carlstrom, J.~E., \& Lay, O.~P. 2001, \apj, 553, 1036
\bibitem[Woody et al.(2000)]{woody} Woody, D.~J., Carpenter, J., \& Scoville, N. 2000, in ASP Conf. Ser. 217, Imaging at Radio Through Submillimeter Wavelengths, ed J.~G. Mangum, \& S.~J.~E. Radford (San Francisco: ASP), 317
\bibitem[Zivanovic et al.(1995)]{zivanovic} Zivanovic, S.~S., Forster, J.~R., \& Welch, W.~J. 1995, Radio Sci, 30, 877
\end{thebibliography}
\end{document}